\documentclass[11pt]{revtex4}
\usepackage{mathtools}
\usepackage{hyperref}
\usepackage{pdflscape}
\usepackage{graphics}
\usepackage{subfigure}
\usepackage{epsfig}
\usepackage{amsmath,amsfonts}
\usepackage{amssymb}
\usepackage{color}
\usepackage{ulem}
\begin{document}
\title{Polarization modes of gravitational waves in generalized Proca theory}

\author{Yu-Qi Dong$^{a,b,c}$}
\email{dongyq21@lzu.edu.cn}

\author{Yu-Qiang Liu$^{a,b,c}$}
\email{liuyq18@lzu.edu.cn}

\author{Yu-Xiao Liu$^{a,b,c}$}
\email{liuyx@lzu.edu.cn, corresponding author}

\affiliation{$^{a}$ Institute of Theoretical Physics \& Research Center of Gravitation, Lanzhou University, Lanzhou 730000, China\\
$^{b}$ Key Laboratory of Quantum Theory and Applications of MoE, Lanzhou University, Lanzhou 730000, China\\
$^{c}$ Lanzhou Center for Theoretical Physics \& Key Laboratory of Theoretical Physics of Gansu Province, Lanzhou University, Lanzhou 730000, China}

\begin{abstract}
\textbf{Abstract:} In this paper, we study polarization modes of gravitational waves in generalized Proca theory in the homogeneous and isotropic Minkowski background. The results show that the polarizations of gravitational waves depend on the parameter space of this gravity theory and can be divided into quite rich cases by parameters. In some parameter space, it only allows two tensor modes, i.e., the $+$ and $\times$ modes. In some parameter space, besides tensor modes, it also allows one scalar mode, or two vector (vector-$x$ and vector-$y$) modes, or both one scalar mode and two vector modes. The scalar mode is a mixture mode of a breathing mode and a longitudinal mode, or just a pure breathing mode.
Interestingly, it is found that the amplitude of the vector modes is related to the speed of the tensor modes. This allows us to give the upper bound of the amplitude of the vector modes by detecting the speed of the tensor modes. {Specifically, if the speed of tensor modes is strictly equal to the speed of light, then the amplitude of vector modes is zero}.
\end{abstract}
	
\maketitle

\section{Introduction}
\label{sec: intro}
The successful detection of gravitational waves \cite{Abbott1,Abbott2,Abbott3,Abbott4,Abbott5} implies the arrival of the gravitational wave astronomy era. In addition to using electromagnetic waves, we can also develop astronomy and cosmology by detecting gravitational waves \cite{Shi Pi,Andrea Addazi,Rong-Gen Cai,Zhi-Chao Zhao,Zong-Kuan Guo}. Not only that, gravitational waves can also help us test various modified gravities and deepen our understanding of gravity \cite{R. Cai2,Ligong Bian}.

Polarization is an important property of gravitational waves. In 1973, Eardley, Lee and Lightman pointed out that there are up to six independent polarization modes of gravitational waves in the general four-dimensional metric theory \cite{Eardley}. For a specific modified gravity, due to the field equations constraining the possible value of the Riemann tensor, it generally does not allow all six polarization modes, but rather allows a subset of them. For example, for general relativity, the vacuum Einstein field equation requires the Ricci tensor is zero, which results in the theory having only two polarization modes of gravitational waves, namely the $+$ mode and the $\times$ mode. Different theories predict different classes of polarization modes, so it can be expected that some of them will be excluded from the detection of polarization modes of gravitational waves in the future.

In many modified gravities, the polarization modes of gravitational waves have been studied, like $f(R)$, Horndeski, Palatini generalized Brans-Dicke, Palatini Horndeski, teleparallel Horndeski, Einstein-aether, tensor-vector-scalar, Horava, scalar-tensor-vector, $f(T)$, dCS, EdGB and Bumblebee theories \cite{f(R,Horndeski0,TeVeS,Horava,STVG,fT,dCS and EdGB,Y.Dong,L.Shao,S. Bahamonde,J.Lu}. {By the way, there are also some studies on the generation of gravitational waves in modified gravities, for example, Refs. \cite{Emmanuele Battista1,Emmanuele Battista2,Zhao Li}.} The ground-based gravitational wave detectors of LIGO, VIRGO and KAGRA are located at different locations on the Earth. They can detect gravitational waves in different two-dimensional planes. Therefore, different detectors can cooperate to detect the polarization modes of gravitational waves. In this regard, some related work has already begun \cite{Abbott4,Abbott5,H. Takeda,B. P . Abbottet al.1,B. P . Abbottet al.2,B. P . Abbottet al.3,Atsushi Nishizawa,Kazuhiro Hayama,Maximiliano Isi1,Maximiliano Isi2,K. Chatziioannou,Hiroki Takeda,Yuki Hagihara,Peter T. H. Pang,B. P . Abbottet al.4}. Polarization information of gravitational waves can also be detected using Pulsar Timing Arrays (PTA) \cite{YiPeng Jing}. For example, in 2021, Huang et al. discovered a tentative indication for scalar transverse gravitational waves in NANOGrav 12.5-year data set \cite{Z. Chen}. Furthermore, one can also expect to use space gravitational wave detectors such as Lisa, Taiji and TianQin {\cite{lisa,taiji,Lisa-taiji,tianqin}} to detect polarization modes of gravitational waves in the future.

Adding additional field is one way to modify gravity. The additional degrees of freedom may explain inflation and the accelerated expansion of the Universe \cite{AA.Starobinsky,A. H. Guth,K.Sato,L.Heisenberg0}. If the additional field is a scalar field, then such a theory is called scalar-tensor theory. In the metric formalism, the most general scalar-tensor theory that can derive second-order field equations is Horndeski theory \cite{Horndeski}. This theory was first discovered by Horndeski in 1974. Later, one realized that extending {the scalar Galileon theory} \cite{A. Nicolis} to curved space-time would lead to the rediscovery of Horndeski theory \cite{C. Deffayet1,C. Deffayet2}.

 The observation of the anisotropy of the cosmic microwave background (CMB) may indicate the existence of a preferred direction in the universe \cite{Planck Collaboration}. This may be naturally generated by an additional vector field \cite{L.Heisenberg0}. Therefore, in addition to adding a scalar field, we can also add an additional vector field to modify gravity. Such a theory is called vector-tensor theory. Similar to the scalar Galileon theory, in 2014, {Tasinato constructed a vector Galileon theory \cite{G. Tasinato1}. Heisenberg further gave generalization of the Proca action in curved space-time} \cite{Heisenberg}. This covariant vector-tensor theory constitutes the Galileon-type generalization of the Proca action, and it is called generalized Proca theory. When one considers the Lagrangian of a vector field with derivative self-interactions with priori arbitrary coefficients, which does not have ghost-like pathologies, and extends it to the case of curved spacetime, generalized Proca theory will be obtained \cite{Heisenberg}. The field equations of generalized Proca theory are second-order, so there is no the Ostrogradsky instability in this theory \cite{M.Ostrogradsky}. There has been a lot of research on this theory, such as cosmology {\cite{Antonio De Felice,Lavinia Heisenberg,Chao Qiang Geng,Lavinia Heisenberg4,Lavinia Heisenberg5}} and black holes \cite{G.Tasinato2,Eugeny Babichev,Lavinia Heisenberg2,Lavinia Heisenberg3,Sebastian Garcia-Saenz}.

In this paper, we will study polarization modes of gravitational waves in generalized Proca theory in the range of a linear analysis. We consider a homogeneous and isotropic Minkowski background and our goal is to obtain the linearized field equations and then analyze the polarization modes of gravitational waves allowed by these equations. Therefore, in Sec. \ref{sec: 2}, we review the action of generalized Proca theory. In Sec. \ref{sec: 3}, we obtain the background equations and linear perturbation equations. In Sec. \ref{sec: 4}, we combine the perturbations into some gauge invariants. These gauge invariants can help us analyze the linear perturbation equations. In Sec. \ref{sec: 5}, we analyze the polarization modes of gravitational waves in generalized Proca theory. {In fact, in Refs. \cite{Antonio De Felice} and \cite{Lavinia Heisenberg5}, the authors analyzed all the permutations and propagating degree of freedom of generalized Proca theory  on top of a Friedmann-Lema\^{i}tre-Robertson-Walker  background, and the perturbation equations and wave speed expressions we need can be directly obtained from these results by taking the Minkowski background limit}.

We will use the natural system of units $c=G=1$ and consider four-dimensional space-time in this paper. We set the metric signature as $(-,+,+,+)$. The Latin alphabet indices $(a,b,c,d,e)$ range over space-time indices ($0,1,2,3$), and the Latin alphabet indices $(i,j,k)$ range over space indices ($1,2,3$) which point to ($+x,+y,+z$) directions respectively.

\section{generalized Proca theory}
\label{sec: 2}
Generalized Proca theory is a vector-tensor theory. In addition to being related to the metric $g_{ab}$, the action also depends on a vector field $A^{a}$.

The action of generalized Proca theory is \cite{Heisenberg,Antonio De Felice,Antonio De Felice2}
\begin{eqnarray}
	\label{action}
	S\left(g_{ab},A^{a}\right) = \int d^4x \sqrt{-g}~
	\Bigl(\mathcal{L}_{F}+\mathcal{L}_{2}+\mathcal{L}_{3}+\mathcal{L}_{4}+\mathcal{L}_{5}\Bigl),
\end{eqnarray}
where
\begin{eqnarray}
	\label{LF}
	\mathcal{L}_{F}&=&-\frac{1}{4}F_{ab}F^{ab},
	\\
	\label{L2}
	\mathcal{L}_{2}&=&G_{2}(X),
    \\
	\label{L3}
	\mathcal{L}_{3}&=&G_{3}(X)\nabla_{a}A^{a},
	\\
	\label{L4}
	\mathcal{L}_{4}&=&G_{4}(X){R}
	                +G_{4,X}(X)\left[
	                               \left(\nabla_{a}A^{a}\right)^{2}
	                               +c_{2}\nabla_{a}A_{b}\nabla^{a}A^{b}
	                               -\left(1+c_{2}\right)\nabla_{a}A_{b}\nabla^{b}A^{a}
	                           \right],
	\\
	\label{L5}
	\nonumber
	\mathcal{L}_{5}&=&G_{5}(X)\left({R}_{ab}-\frac{1}{2}{g}_{ab}{R}\right)\nabla^{a}A^{b}
	\\ \nonumber
	&-&\frac{1}{6}G_{5,X}(X)
	\left[
	  \left(\nabla_{a}A^{a}\right)^{3}
	  -3d_{2}\nabla_{a}A^{a}\nabla_{b}A_{c}\nabla^{b}A^{c}
	  -3\left(1-d_{2}\right)\nabla_{a}A^{a}\nabla_{b}A_{c}\nabla^{c}A^{b}
	\right.\nonumber
	\\
	&+&
	\left.
	\left(2-3d_{2}\right)\nabla_{a}A_{b}\nabla^{c}A^{a}\nabla^{b}A_{c}
	+3d_{2}\nabla_{a}A_{b}\nabla^{c}A^{a}\nabla_{c}A^{b}
	\right].
\end{eqnarray}
Here, $F_{ab}=\nabla_{a}A_{b}-\nabla_{b}A_{a}$, $R_{ab}$ is the Ricci tensor and $R$ is the Ricci scalar, $c_{2},d_{2}$ are constants, $G_{2},G_{3},G_{4}$ and $G_{5}$ are arbitrary functions of the variable $X=-\frac{1}{2}g_{ab}A^{a}A^{b}$. In addition, $G_{n,X}={dG_{n}}/{dX}$ ($n=2,3,4,5$). The remaining derivatives are still represented by this notation. For example, $G_{2,{XX}}$ represents the second-order derivative of $G_{2}$ with respect to the variable $X$. In the action of generalized Proca theory, $L_3$, $L_4$, and $L_5$ correspond to the first, second, and third power terms of the derivative interaction of the vector field, respectively. And $G_{4}(X){R}$ and $G_{5}(X)\left({R}_{ab}-\frac{1}{2}{g}_{ab}{R}\right)\nabla^{a}A^{b}$ are the non-minimal couplings between the vector field and curvature that must be introduced in order for the field equation to be second-order. It can be seen that unlike Einstein-aether theory \cite{Christopher Eling}, the action of generalized Proca theory does not have a Lagrangian multiplier for a priori fixed norm of the vector field.

The first term of $\mathcal{L}_{4}$ can be equivalently written as $(16\pi)^{-1}R+\left[G_{4}(X)-(16\pi)^{-1}\right]R$. Therefore, it is easy to see that this theory includes the Einstein-Hilbert action. We hope that generalized Proca theory can retain the contribution of the Einstein-Hilbert term, so it requires
\begin{eqnarray}
\label{G4neq0}
G_{4}(X) \neq 0.
\end{eqnarray}

\section{Background equations and linear perturbation equations}
\label{sec: 3}
Now consider a homogeneous and isotropic Minkowski background
\begin{eqnarray}
	\label{background}
	g_{ab}=\eta_{ab},\quad A^a=\mathring{A}^{a}=\left(A,0,0,0\right),
\end{eqnarray}
where $\mathring{A}^{a}$ is a constant vector and $\eta_{ab}$ is the Minkowski metric. Due to spatial isotropy, the vector $\mathring{A}^{a}$ only has one nonvanishing temporal component $A$, and its spatial components are all zero.

By substituting the assumption (\ref{background}) into the field equations derived from the variation of the action (\ref{action}) with respect to the metric tensor $g^{ab}$ and the vector field $A^{a}$, respectively, we can determine that $\mathring{A}^{a}$ should satisfy the following background equations:
\begin{eqnarray}
	\label{background equation tensor}
	-\frac{1}{2}\mathring{G}_{2}\eta_{ab}+\frac{1}{2}\mathring{G}_{2,X}\mathring{A}_{a}\mathring{A}_{b}
       &=&0,
	\\
	\label{background equation vector}
	-\mathring{G}_{2,X}\mathring{A}_{a}&=&0.
\end{eqnarray}
Here and below, the notation ``$\circ$" above the letter means that the corresponding function takes the value of $X = -\frac{1}{2}\eta_{ab}\mathring{A}^{a}\mathring{A}^{b}$. Using the background equations (\ref{background equation tensor}) and (\ref{background equation vector}), we can divide the background solution (\ref{background}) into two cases:

\begin{center}
	\begin{tabular}{l}
{Case A}:\quad $\mathring{G}_{2}=0,~\mathring{G}_{2,X} \neq 0,~A=0$.
\\
{Case B}:\quad $\mathring{G}_{2}=0,~\mathring{G}_{2,X}=0$,~$A=\text{constant}$.
    \end{tabular}
\end{center}

To describe gravitational waves, we investigate perturbations of the background
\begin{eqnarray}
	\label{perturbations}
	g_{ab}=\eta_{ab}+h_{ab},\quad A^{a}=\mathring{A}^{a}+B^{a};\quad
	\mid h_{ab} \mid\!\!\! ~ \ll\!\!\! ~ 1,
	\quad \mid B^{a} \mid \!\!\! ~\ll \!\!\! ~\mid \mathring{A}^{a} \mid.
\end{eqnarray}
In the following, we use $\eta_{ab}$ and $\eta^{ab}$ to lower and raise the space-time index and define $h=\eta^{ab}h_{ab}$.

By substituting the perturbations (\ref{perturbations}) into the field equations derived from the variation of the action (\ref{action}), and writing the first-order terms of the perturbations, we can obtain two linear perturbation equations that describes gravitational waves.

Substituting the perturbations into the field equation derived from the variation of vector $A^{a}$, we obtain the first linear perturbation equation
\begin{eqnarray}
	\label{vector perturbations equation}
	\mathcal{V}_{a}=0.
\end{eqnarray}
Here, the specific expression of $\mathcal{V}_{a}$ is very lengthy, and we place it in Appendix \ref{app: A}.

Substituting the perturbations into the field equation derived from the variation of tensor $g^{ab}$, we have the second linear perturbation equation
\begin{eqnarray}
	\label{tensor perturbations equation}
	  \mathcal{T}_{ab}=0.
\end{eqnarray}
Here, we also list the expression of $\mathcal{T}_{ab}$ in Appendix \ref{app: A}.

It should be noted that for the case of $A=0$, i.e., $\mathring{A}^{a}=0$, it is easy to analyze  polarization modes of gravitational waves. In this case, Eq. (\ref{tensor perturbations equation}) becomes
\begin{eqnarray}
	\label{tensor perturbations equation A=0}
	-2\mathring{G}_{2}h_{ab}
	-2\mathring{G}_{4}
	\left(
	\partial_{a}\partial_{b}h
	-\partial_{a}\partial_{c}h^{c}_{~b}
	-\partial_{b}\partial_{c}h^{c}_{~a}
	+\Box h_{ab}
	+\eta_{ab} \partial_{c}\partial_{d}h^{cd}
	-\eta_{ab} \Box h
	\right)
	=0.
\end{eqnarray}
On the other hand, we always have $\mathring{G}_{2}=0$ for the background solution (\ref{background}). In addition, since $G_{4}(X) \neq 0$, Eq. (\ref{tensor perturbations equation A=0}) can be further written as
\begin{eqnarray}
	\label{R=0}
	\mathop{G}^{(1)}~\!\!\!_{ab}
	=
	-\frac{1}{2}
	\left(
	\partial_{a}\partial_{b}h
	-\partial_{a}\partial_{c}h^{c}_{~b}
	-\partial_{b}\partial_{c}h^{c}_{~a}
	+\Box h_{ab}
	+\eta_{ab} \partial_{c}\partial_{d}h^{cd}
	-\eta_{ab} \Box h
	\right)
	=0.
\end{eqnarray}
Here, $G_{ab}$ is the Einstein tensor, and the notation ``$(1)$" above the letter $G$ means that only the first-order term of the perturbations is taken here. It can be seen that under linear approximation, Eq. (\ref{R=0}) is no different from Einstein's field equation. Therefore, just like  general relativity, there are only two tensor modes ($+$ and $\times$) propagating at the speed of light for the case of $A=0$. It should be pointed out that this does not mean that, when $A=0$, general Proca theory will return to general relativity. When we consider not only linear-order but also the full nonlinear field equation, we will find that the field equations of generalized Proca theory contain higher-order terms of vector field perturbations.

One may ask whether another linear perturbation equation (\ref{vector perturbations equation}) further constrains the tensor mode gravitational waves. The answer is no. Note that only the transverse traceless part $h^{TT}_{ij}$ of the metric perturbation $h_{ab}$ contributes to the tensor mode gravitational waves, and Eq. (\ref{vector perturbations equation}) is a vector type equation with a single index. We can know that this equation does not constrain the value of $h^{TT}_{ij}$,  and therefore does not constrain the tensor mode gravitational waves.

Through the above analysis, we find that when the background solution (\ref{background}) belongs to Case A, it only allows two tensor ($+$ and $\times$) modes propagating at the speed of light. The result is the same for Case B with $A=0$. Therefore, in the following sections, we only need to analyze the case where the background solution (\ref{background}) meets the following conditions:
\begin{eqnarray}
	\label{case B A neq 0}
 \mathring{G}_{2}=0,~\mathring{G}_{2,X}=0,~A \neq 0.
\end{eqnarray}

\section{gauge invariants}
\label{sec: 4}
In this section, we will introduce a method for simplifying linear perturbation equations. After using this method, the linear perturbation equations can be decomposed into independent tensor, vector, and scalar equations. These equations can be represented by some gauge invariants combined with perturbations.

Under spatial rotation transformation, $h_{ij}$ is transformed like a second-order tensor, $h_{0i}$,  $B^{i}$ are transformed like vectors, and $h_{00}$, $B^{0}$ are transformed like scalars. We can uniquely decompose the perturbations as follows \cite{R. Jackiw,Eanna E Flanagan,L.Shao}:
\begin{eqnarray}
	\label{decompose perturbations}
     B^{0}&=&B^{0},\nonumber \\
	 B^{i}&=&\partial^{i}\omega+\mu^{i},\nonumber \\
     h_{00}&=&h_{00}, \\
     h_{0i}&=&\partial_{i}\gamma+\beta_{i},\nonumber\\
     h_{ij}&=&h^{TT}_{ij}+\partial_{i}\epsilon_{j}+\partial_{j}\epsilon_{i}
      +\frac{1}{3}\delta_{ij}H+(\partial_{i}\partial_{j}-\frac{1}{3}\delta_{ij}\Delta)\zeta. \nonumber
\end{eqnarray}
Here,
\begin{eqnarray}
     &\partial_{i}\mu^{i}=\partial_{i}\beta^{i}=\partial_{i}\epsilon^{i}=0,  \\
     &\delta^{ij}h^{TT}_{ij}=\partial^{i}h^{TT}_{ij}=0.
\end{eqnarray}
Here and below, we use $\delta_{ij}$ and $\delta^{ij}$ to lower and raise the space index. As can be seen, Eq. (\ref{decompose perturbations}) uniquely decomposes a spatial vector into a scalar part and a transverse vector part, and uniquely decomposes a spatial tensor into two scalar parts, a transverse vector part, and a transverse traceless tensor part.

Considering that the left sides of Eqs.~(\ref{vector perturbations equation}) and (\ref{tensor perturbations equation}) are actually a vector and a tensor, respectively, we can also apply this decomposition method to the linear perturbation equations. Due to the uniqueness of the decomposition, requiring the equations to be zero is equivalent to requiring that each decomposed part of the linear perturbation equations be zero, so we can obtain a series of independent tensor, vector and scalar equations. Due to the spatial homogeneous and isotropic of the background solution (\ref{background}), by substituting (\ref{decompose perturbations}) into Eqs.~(\ref{vector perturbations equation}) and (\ref{tensor perturbations equation}), we can see that these decomposed scalar, vector and tensor equations only rely on tensor, vector and scalar perturbations, respectively \cite{Weinberg}. Therefore, the linear perturbation equations are decoupled.

We can also continue to combine the perturbations in these equations into gauge invariants. Under linear approximation, the gauge transformations of the metric and the vector field are \cite{Robert Bluhm}
\begin{eqnarray}
	\label{gauge transformation}
	h_{ab} &\rightarrow& h_{ab}-\partial_{a}\xi_{b}-\partial_{b}\xi_{a}, \\
	B^{a} &\rightarrow& B^{a}+\mathring{A}^{b}\partial_{b}\xi^{a}.
\end{eqnarray}
Here, $\xi^{a}$ is an arbitrary function of space-time coordinates.

Under the above gauge transformations, the spatial tensor
\begin{eqnarray}
	\label{tensor gauge invariant}
	h^{TT}_{ij},
\end{eqnarray}
the two spatial vectors
\begin{eqnarray}
	\label{vector gauge invariant}
\begin{array}{l}
  \Xi_{i}=\beta_{i}-\partial_{0}\epsilon_{i},  \\
	\Sigma_{i}=\mu_{i}+A\partial_{0}\epsilon_{i},
\end{array}	
\end{eqnarray}
and the four spatial scalars
\begin{eqnarray}	
\begin{array}{l}
  \phi \,= -\frac{1}{2}h_{00}+\partial_{0}\gamma-\frac{1}{2}\partial_{0}\partial_{0}\zeta,  \\
	\Theta = \frac{1}{3}\left(H-\Delta\zeta\right),  \\
	\Omega= B^{0}-A\partial_{0}\gamma+\frac{1}{2}A\partial_{0}\partial_{0}\zeta,  \\
	\Psi = \omega+\frac{1}{2}A\partial_{0}\zeta,
\end{array} \label{scalar gauge invariant}
\end{eqnarray}
are gauge invariants \cite{Eanna E Flanagan,L.Shao}.
In the next section, we will check that the scalar, vector, and tensor equations decomposed from Eqs.~(\ref{vector perturbations equation}) and (\ref{tensor perturbations equation}) can be represented by the gauge invariants (\ref{tensor gauge invariant})-(\ref{scalar gauge invariant}) combined with perturbations.

\section{Polarization modes of gravitational waves}
\label{sec: 5}
In this section, we will analyze the polarization modes of gravitational waves in generalized Proca theory in Case B.

We detect gravitational waves by detecting the relative displacement of free particles \cite{MTW}. In this paper, similar to Ref. \cite{Eardley}, we assume that the matter field only has minimal coupling with the metric $g_{ab}$, while the vector field $A^{a}$ is not directly coupled with the matter field. So, the action describing the motion of a free particle is
\begin{eqnarray}
	\label{a free particle}
	S=\int ds=\int \sqrt{|g_{ab}dx^{a}dx^{b}|},
\end{eqnarray}
the relative motion between two adjacent free test particles used to detect the polarization modes of gravitational waves satisfies the equation of geodesic deviation \cite{MTW}:
\begin{eqnarray}
	\label{equation of geodesic deviation}
	\frac{d^{2}\eta_{i}}{dt^{2}}=-\mathop{R}^{(1)}\!_{i0j0}\eta^{j}.
\end{eqnarray}
Here, $\eta_{i}$ is the relative displacement of the two test particles and $R_{i0j0}$ is the $(i0j0)$ component of the Riemannian tensor. The notation ``$(1)$" above the letter $R$ means that only the first-order term of the perturbations is taken here. It can be seen from Eq. (\ref*{equation of geodesic deviation}) that $R_{i0j0}$ will completely determine the motion behavior of the  free test particles under the gravitational waves, and the polarization modes of gravitational waves are defined by different values of $R_{i0j0}$ \cite{MTW}.

Using Eqs. (\ref{decompose perturbations}), (\ref{tensor gauge invariant})-(\ref{scalar gauge invariant}), we can write $R_{i0j0}$ as
\begin{eqnarray}
	\label{Ri0j0 gauge invariant}
	\mathop{R}^{(1)}\!_{i0j0}=-\frac{1}{2}\partial_{0}\partial_{0}h^{TT}_{ij}
                             +\frac{1}{2}\partial_{0}\partial_{i}\Xi_{j}
                             +\frac{1}{2}\partial_{0}\partial_{j}\Xi_{i}
                             +\partial_{i}\partial_{j}\phi
                             -\frac{1}{2}\delta_{ij}\partial_{0}\partial_{0}\Theta.
\end{eqnarray}
It can be seen that among several gauge invariants, only $h^{TT}_{ij},\Xi_{i},\phi,\Theta$ contribute to gravitational waves, while $\Sigma_{i},\Omega,\Psi$ do not cause the relative displacement of the test particles.

Now taking the propagation direction of gravitational waves as $+z$ direction. Mark the
components of $R_{i0j0}$ as \cite{Eardley,Y.Dong}
\begin{eqnarray}
	\label{P1-P6}
	R_{i0j0}=\begin{pmatrix}
		P_{4}+P_{6} & P_{5} & P_{2}\\
		P_{5}       & -P_{4}+P_{6}  & P_{3}\\
		P_{2}       &  P_{3}   &   P_{1}
	\end{pmatrix}.
\end{eqnarray}
We can define six independent polarization modes of gravitational waves $P_{1},\cdots,P_{6}$. Any plane gravitational wave can be written as a linear combination of these six modes. We show the relative motions of the test particles under these six polarization modes in Fig. 1.

\begin{figure*}[htbp]
	\makebox[\textwidth][c]{\includegraphics[width=1.2\textwidth]{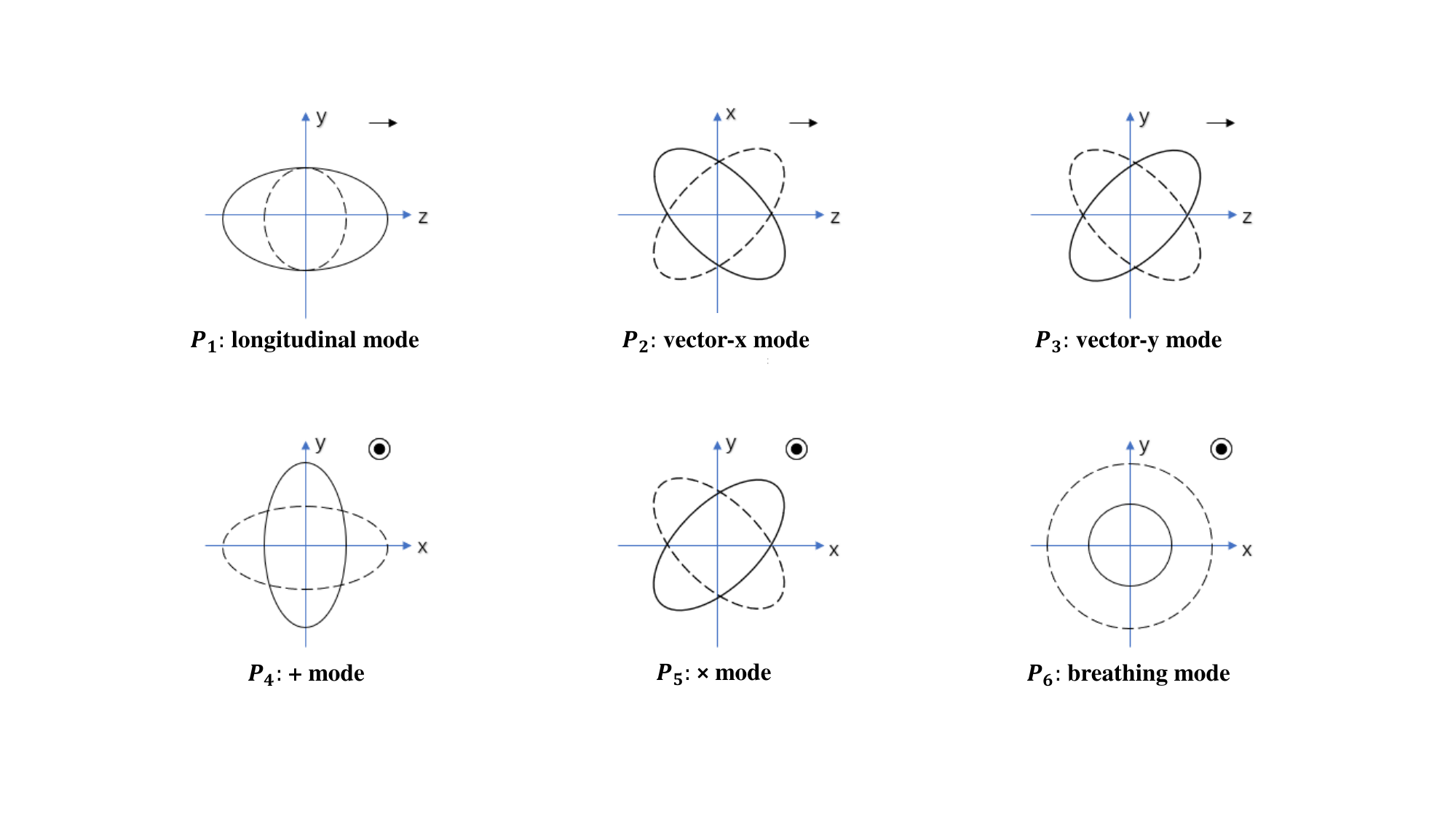}}
	\caption{Six polarization modes of gravitational waves \cite{Eardley}. We have assumed that the gravitational waves propagate in the $+z$ direction. The solid and dotted lines are the cases of a circle of test particles when the phases of waves are $0$ and $\pi$, respectively. There is no relative motion between test particles in the direction of the third axis that is not drawn.}
	\label{fig: 1}
\end{figure*}

Combining Eqs.~(\ref{Ri0j0 gauge invariant}) and (\ref{P1-P6}), we can obtain the expressions of the six polarization modes
\begin{eqnarray}
	\label{P1-P6 gauge invariant}
\begin{array}{l}
  P_{1}=\partial_{3}\partial_{3}\phi-\frac{1}{2}\partial_{0}\partial_{0}\Theta, \quad
	P_{2}=\frac{1}{2}\partial_{0}\partial_{3}\Xi_{1},\\
  P_{3}=\frac{1}{2}\partial_{0}\partial_{3}\Xi_{2},  \quad \quad\quad\quad\,\,
  P_{4}=-\frac{1}{2}\partial_{0}\partial_{0}h^{TT}_{11}, \\
	P_{5}=-\frac{1}{2}\partial_{0}\partial_{0}h^{TT}_{12}, \quad\quad~~~
	P_{6}=-\frac{1}{2}\partial_{0}\partial_{0}\Theta.
\end{array}
\end{eqnarray}
It should be pointed out that in this calculation, we assume that the propagation direction of gravitational waves is in the $+z$ direction. Therefore, when we consider wave solutions, $h^{TT}_{ij}, \Xi_{i}, \phi, \Theta$ are functions of only $t$ and $z$. This will result in, for example, $h^ {TT}_{22}=-h^{TT}_{11} $ and $\partial_{1}\partial_{1}\phi=0$. It can be seen that $h^{TT}_{ij}$ only contributes to tensor ($+$ and $\times$) modes, $\Xi_{i}$ only contributes to vector (vector-x and vector-y) modes, $\phi$ only contributes to longitudinal mode, and $\Theta$ contributes to both breathing mode and longitudinal mode.

Now, we will use the method introduced in Sec. \ref{sec: 4} to simplify the linear perturbation equations (\ref{vector perturbations equation}) and (\ref{tensor perturbations equation}) and analyze the polarization modes of gravitational waves in generalized Proca theory in case (\ref{case B A neq 0}).

 \subsection{Tensor mode}
 \label{tensor}
The equation describing the tensor mode gravitational waves can be derived from the transverse traceless tensor part of the $ij$ component in Eq. (\ref{tensor perturbations equation}), which is
\begin{eqnarray}
	\label{tensor mode equation}
	\left(\mathring{G}_{4}-\mathring{G}_{4,X}{A}^2\right)\partial_{0}\partial_{0}h^{TT}_{ij}
	-\mathring{G}_{4}\Delta h^{TT}_{ij}
	=0.
\end{eqnarray}
{This equation is consistent with the result of taking Minkowski limit in Refs. \cite{Antonio De Felice,Lavinia Heisenberg5}.} In order for generalized Proca theory to have tensor mode ($+$ and $\times$) gravitational waves, combining Eqs. (\ref{G4neq0}) and (\ref{tensor mode equation}), we should have the condition
\begin{eqnarray}
	\label{G4-G4XA2 neq 0}
	\mathring{G}_{4}-\mathring{G}_{4,X}{A}^2
	\neq 0.
\end{eqnarray}
If the condition (\ref{G4-G4XA2 neq 0}) is not satisfied, then Eq. (\ref{tensor mode equation}) will become $\Delta h^{TT}_{ij}=0$. It indicates that $h^{TT}_{ij}$ has no wave solution, so this is not what we expected.

With the condition (\ref{G4-G4XA2 neq 0}), we can obtain from Eq. (\ref{tensor mode equation}) that the speed of the tensor mode gravitational waves $c_{T}$ is
\begin{eqnarray}
	\label{cT}
	{c_{T}}^2=\frac{\mathring{G}_{4}}{\mathring{G}_{4}-\mathring{G}_{4,X}{A}^2}.
\end{eqnarray}
If ${c_{T}}^2 \textless 0$, the solution will exponentially diverge, and then it is linearly unstable. To ensure the linear stability of the solution, we require ${c_{T}}^2 \textgreater 0$.

GW170817 and GRB170817A require the speed of tensor mode gravitational waves $c_T$ to meet \cite{B.P.Abbott000,B.P.Abbott111}
\begin{eqnarray} 	
	\label{GW170817}
	-3\times10^{-15} \leq c_T-1 \leq 7\times10^{-16}.
\end{eqnarray}
This result will constrain the values of the theoretical parameters. Considering that Eq. (\ref{GW170817}) requires $c_{T}$ not to deviate significantly from the speed of light. Using Eq. (\ref{cT}), it shows that
\begin{eqnarray}
	\mid\mathring{G}_{4,X}{A}^2 \mid \!\!\! ~\ll \!\!\! ~\mid \mathring{G}_{4} \mid.
\end{eqnarray}
Therefore, we can expand $c_{T}^2$ as
\begin{eqnarray}
		\label{cT2=1+d}
	c_{T}^2 \approx 1+\frac{\mathring{G}_{4,X}{A}^2}{\mathring{G}_{4}}.
\end{eqnarray}
Accordingly, $c_{T}$ has an approximation of
\begin{eqnarray}
	c_{T} \approx 1+\frac{\mathring{G}_{4,X}{A}^2}{2\mathring{G}_{4}}.
\end{eqnarray}
Therefore, it can be seen that the constrain (\ref{GW170817}) requires
\begin{eqnarray}
	\label{G4-G4XA2/G4 less 10-15}
	\left | \frac{\mathring{G}_{4,X}{A}^2}{\mathring{G}_{4}}\right |
       \!\! ~\lesssim \!\! ~ 10^{-15}.
\end{eqnarray}
In particular, when the speed of tensor mode gravitational waves is strictly equal to the speed of light, we have $\mathring{G}_{4,X}=0$.

 \subsection{Vector mode}
 \label{vector}
We can derive three equations describing vector mode gravitational waves from the linear perturbation equations (\ref{vector perturbations equation}) and (\ref{tensor perturbations equation}).

The first equation can be derived from the transverse vector part of the $i$ component of Eq. (\ref{vector perturbations equation}), which is
\begin{eqnarray}
	\label{vector mode equation 1}
	\left(1-2c_{2}\mathring{G}_{4,X}\right) \Box \left(\Sigma_{i}+A\Xi_{i}\right)
	-\mathring{G}_{4,X}A\Delta\Xi_{i}
	=0.	
\end{eqnarray}
The second and third equations can be derived respectively from the transverse vector parts of the $ij$ and $0i$ components of Eq. (\ref{tensor perturbations equation}):
\begin{eqnarray}
	\label{vector mode equation 2}
	\mathring{G}_{4}\Xi_{i}
	+\mathring{G}_{4,X}A\Sigma_{i}
	=0, \\
	\label{vector mode equation 3}
	\left(1-2c_{2}\mathring{G}_{4,X}\right) A \Box \left(\Sigma_{i}+A\Xi_{i}\right)
	-\mathring{G}_{4}\Delta \Xi_{i}
	-\mathring{G}_{4,X}A\Delta \left(\Sigma_{i}+A\Xi_{i}\right)
	=0.
\end{eqnarray}
{These equations are consistent with the results of taking Minkowski limit in Refs. \cite{Antonio De Felice,Lavinia Heisenberg5}.} Perhaps one would ask why there are only two variables $\Xi_{i}$, $\Sigma_{i}$ but three equations here. In fact, these three equations are not independent of each other. We can easily find that Eq. (\ref{vector mode equation 3}) can be obtained using Eqs. (\ref{vector mode equation 1}) and (\ref{vector mode equation 2}). Therefore, we only need to consider the first two.

Using Eq. (\ref{vector mode equation 2}), we have
\begin{eqnarray}
	\label{Xi=Sigma}
	\Xi_{i}=-\frac{\mathring{G}_{4,X}A}{\mathring{G}_{4}}\Sigma_{i}.
\end{eqnarray}
It can be seen that when $\mathring{G}_{4,X}=0$, we have $\Xi_{i}=0$. However, from Eq. (\ref{Ri0j0 gauge invariant}), it can be seen that the gauge invariant vector that contributes to gravitational waves is only $\Xi_{i}$. Therefore, in this case, generalized Proca theory does not allow vector mode gravitational waves.

When $\mathring{G}_{4,X} \neq 0$, we can substitute Eq. (\ref{Xi=Sigma}) into Eq. (\ref{vector mode equation 1}), and obtain
\begin{eqnarray}
	\label{vector mode wave equation}
	\mathcal{M}\partial_{0}\partial_{0}\Sigma_{i}
    -\mathcal{N}\Delta\Sigma_{i}=0.
\end{eqnarray}
where
\begin{eqnarray}
	\mathcal{M}&=& \left(1-2c_{2}\mathring{G}_{4,X}\right)\left(\mathring{G}_{4}-\mathring{G}_{4,X}A^2\right),
	 \\
    \mathcal{N} &=& \mathcal{M} + {\mathring{G}_{4,X}}^2 A^2.
\end{eqnarray}
This is an equation for the variable $\Sigma_{i}$. When we solve the wave solution of $\Sigma_{i}$ from Eq. (\ref{vector mode wave equation}), we can obtain $\Xi_{i}$ with Eq. (\ref{Xi=Sigma}), and then obtain the wave solution of vector mode gravitational waves. We already have the condition (\ref{G4-G4XA2 neq 0}), so when considering $\mathring{G}_{4,X} \neq 0$, we can study this equation in the following three cases:

\textbf{Case 1}: When $1-2c_{2}\mathring{G}_{4,X}=0$, Eq. (\ref{vector mode wave equation}) becomes $\Delta\Sigma_{i}=0$. It indicates that $\Sigma_{i}$ has no wave solution, so we have $\Xi_{i}=\Sigma_{i}=0$. This case does not allow vector mode gravitational waves.

\textbf{Case 2}: When $\left(1-2c_{2}\mathring{G}_{4,X}\right)\left(\mathring{G}_{4}-\mathring{G}_{4,X}A^2\right)+{\mathring{G}_{4,X}}^2 A^2=0$, Eq. (\ref{vector mode wave equation}) becomes $\partial_{0}\partial_{0}\Sigma_{i}=0$. It also indicates that $\Sigma_{i}$ has no wave solution, so we have $\Xi_{i}=\Sigma_{i}=0$. This case also does not allow vector mode gravitational waves.

\textbf{Case 3}: When $1-2c_{2}\mathring{G}_{4,X} \neq 0$ and $\left(1-2c_{2}\mathring{G}_{4,X}\right)\left(\mathring{G}_{4}-\mathring{G}_{4,X}A^2\right)+{\mathring{G}_{4,X}}^2 A^2 \neq 0$, Eq. (\ref{vector mode wave equation}) will have wave solutions of $\Sigma_{i}$. In this case, generalized Proca theory allows two vector (vector-x and vector-y) modes. Using Eq. (\ref{vector mode wave equation}), we obtain the speed of vector mode gravitational waves $c_{V}$:
\begin{eqnarray}
	\label{cV}
	c_{V}^2=1+\frac{{\mathring{G}_{4,X}}^2 A^2}{\left(1-2c_{2}\mathring{G}_{4,X}\right)\left(\mathring{G}_{4}-\mathring{G}_{4,X}A^2\right)}.
\end{eqnarray}
Similar to the previous analysis of tensor modes, in order for the solution to have linear stability, we require $c_{V}^2 \textgreater 0$. It can be seen that in Case 3, vector mode gravitational waves do not propagate at the speed of light, i.e. $c_{V} \neq 1$.

Based on the above research, we find that the existence of vector mode gravitational waves in generalized Proca theory depends on the parameter space. We summarize the results in Table \ref{tab:vector}.

\begin{center}
	\begin{table}[htbp]
		\resizebox{\textwidth}{19.5mm}{
			\begin{tabular}{|c|c|c|}
				\hline\hline
				\textbf{Cases} & \textbf{Conditions} &  \textbf{Vector DoF} \\
				\hline
				case 0 & $\mathring{G}_{4,X} = 0.$ & 0 \\
				\hline
				case 1 &$\mathring{G}_{4,X}\neq 0,~1-2c_{2}\mathring{G}_{4,X}=0.$ & 0 \\
				\hline
				case 2 &$\mathring{G}_{4,X}\neq0,~ (1-2c_{2}\mathring{G}_{4,X})(\mathring{G}_{4}-\mathring{G}_{4,X}A^2)+{\mathring{G}_{4,X}}^2 A^2=0.$ & 0 \\
				\hline
				case 3 &$\mathring{G}_{4,X}\neq0,~ 1-2c_{2}\mathring{G}_{4,X} \neq 0,~ (1-2c_{2}\mathring{G}_{4,X})(\mathring{G}_{4}-\mathring{G}_{4,X}A^2)+{\mathring{G}_{4,X}}^2 A^2 \neq 0.$ & 2 \\
				\hline\hline
		\end{tabular}}
		\caption{The number of vector mode gravitational waves under various cases. The numbers in the rightmost column of the table represent the degrees of freedom of the vector mode gravitational waves in the corresponding case.}
		\label{tab:vector}
	\end{table}
\end{center}

Finally, note that $\Xi_{i}$ in Eq. (\ref{Xi=Sigma}) depends on $\mathring{G}_{4,X}$, which allows us to use the speed of tensor mode gravitational waves to constrain the upper limit of the amplitude of vector mode gravitational waves.

First, all the perturbations are weak relative to the background:
\begin{eqnarray}
	\label{h<<1,B<<A}
	\mid h_{ab} \mid\!\!\! ~ \ll\!\!\! ~ 1,
	 \quad \mid B^{a} \mid \!\!\! ~\ll \!\!\! ~\mid \mathring{A}^{a} \mid.
\end{eqnarray}
Then with Eqs. (\ref{vector gauge invariant}) and (\ref{decompose perturbations}), we have
\begin{eqnarray}
	\label{Sigma<<A}
	\mid \Sigma_{i}  \mid  \!\!\! ~ \ll  \!\!\! ~ \mid A \mid.
\end{eqnarray}
Therefore, using Eqs. (\ref{Xi=Sigma}) and (\ref{cT2=1+d}), the condition (\ref{Sigma<<A}) implies
\begin{eqnarray}
	\label{Xi<<cT}
	\mid \Xi_{i} \mid \!\!\! ~
	= \!\!\! ~\left| \frac{\mathring{G}_{4,X}A}{\mathring{G}_{4}}\Sigma_{i} \right| \!\!\! ~
	\ll \!\!\! ~\left| \frac{\mathring{G}_{4,X}A^2}{\mathring{G}_{4}} \right| \!\!\! ~
	\approx \!\!\! ~\mid c_{T}^2-1 \mid.
\end{eqnarray}
The above inequality gives the relationship between the vector mode amplitude and the tensor mode speed. Then, considering GW170817 and combining with the condition (\ref{G4-G4XA2/G4 less 10-15}), we have from Eq. (\ref{Xi<<cT}) that
\begin{eqnarray}
	\label{Xi<<10-15}
	\mid \Xi_{i} \mid \!\!\!~ \ll \!\!\! ~10^{-15}.
\end{eqnarray}
Using Eqs. (\ref{decompose perturbations}) and (\ref{vector gauge invariant}), the above constrain indicates that the metric perturbation (abbreviated as $h$) corresponding to the vector gravitational waves satisfy $|h| \ll 10^{-15}$. If we further require the speed of tensor modes to be the speed of light, i.e. $c_{T}=1$, then according to Eq. (\ref{Xi<<cT}), we have $\Xi_{i}=0$. At this point, the amplitude of the vector modes is zero, and we can not detect the vector gravitational waves.

 \subsection{Scalar mode}
 \label{scalar}
We can derive six equations describing scalar mode gravitational waves from the linear perturbation equations (\ref{vector perturbations equation}) and (\ref{tensor perturbations equation}).

The first and second equations can be derived from the $0$ component and the scalar part of the $i$ component of Eq. (\ref{vector perturbations equation}), respectively, which are given by
\begin{eqnarray}
	\label{scalar mode equation 1}
	&-2\left(1-2c_{2}\mathring{G}_{4,X}\right)\Delta\left(\Omega+2A\phi+\partial_{0}\Psi\right)
	+\mathring{G}_{3,X}A\left(2\Delta\Psi+3A\partial_{0}\Theta\right)
	\nonumber \\
	&-4\mathring{G}_{4,X}A\Delta{\Theta}
	+2\mathring{G}_{2,XX}A^2\left(\Omega+A\phi\right)
	=0, \\
	\label{scalar mode equation 2}
	&-\left(1-2c_{2}\mathring{G}_{4,X}\right)\partial_{0}\left(\Omega+2A\phi+\partial_{0}\Psi\right)
	-\mathring{G}_{3,X}A\left(\Omega+A\phi\right)
	-2\mathring{G}_{4,X}A\partial_{0}\Theta
	=0.
\end{eqnarray}
The third and fourth {ones can be obtained respectively} from the trace part and the $\partial_{i}\partial_{j}$ part of the $ij$ {components} of Eq. (\ref{tensor perturbations equation}):
\begin{eqnarray}
	\label{scalar mode equation 3}
	&\mathring{G}_{4}\left(-4\Delta\phi-2\Delta\Theta+6\partial_{0}\partial_{0}\Theta\right)
	-3\mathring{G}_{3,X}A^{2}\partial_{0}\left(\Omega+A\phi\right)
	\nonumber \\
	&-4\mathring{G}_{4,X}A\Delta\left(\Omega+A\phi+\partial_{0}\Psi\right)
	-6\mathring{G}_{4,X}A^{2}\partial_{0}\partial_{0}\Theta
	=0,\\
	\label{scalar mode equation 4}
    &\mathring{G}_{4}\left(2\phi+\Theta\right)
    +2\mathring{G}_{4,X}A\left(\Omega+A\phi+\partial_{0}\Psi\right)
    =0.
\end{eqnarray}
Finally, the fifth and sixth equations can be derived respectively from the scalar part of the $0i$ component and the $00$ component of Eq. (\ref{tensor perturbations equation}):
\begin{eqnarray}
	\label{scalar mode equation 5}
	&\left(1-2c_{2}\mathring{G}_{4,X}\right)A\left(\Omega+2A\phi+\partial_{0}\Psi\right)
	+2\mathring{G}_{4}\Theta
	=0,\\
	\label{scalar mode equation 6}
	&4\left(1-2c_{2}\mathring{G}_{4,X}\right)A\Delta\left(\Omega+2A\phi+\partial_{0}\Psi\right)
	+4\mathring{G}_{4}\Delta\Theta
	+4\mathring{G}_{4,X}A^2\Delta\Theta
	\nonumber \\
	&-\mathring{G}_{3,X}A^2\left(2\Delta\Psi+3A\partial_{0}\Theta\right)
	-2\mathring{G}_{2,XX}A^3\left(\Omega+A\phi\right)
	=0.
\end{eqnarray}
{These equations are consistent with the results of taking Minkowski limit in Refs. \cite{Antonio De Felice,Lavinia Heisenberg5}.} The above six equations are not independent of each other. Equation (\ref{scalar mode equation 6}) can be obtained from Eqs. (\ref{scalar mode equation 1}) and (\ref{scalar mode equation 5}), and Equation (\ref{scalar mode equation 2}) {from} Eqs. (\ref{scalar mode equation 3}) and (\ref{scalar mode equation 4})(\ref{scalar mode equation 5}). Therefore, only Eqs. (\ref{scalar mode equation 1}) and (\ref{scalar mode equation 3})-(\ref{scalar mode equation 5}) are independent. Now, we have four equations and four variables $\phi,\Theta,\Psi,\Omega$.

We can use Eqs. (\ref{scalar mode equation 3}) and (\ref{scalar mode equation 4}) to obtain the following equation:
\begin{eqnarray}
	\label{scalar mode equation new}
	2\left(\mathring{G}_{4}-\mathring{G}_{4,X}A^2\right)\partial_{0}\Theta
	-\mathring{G}_{3,X}A^{2}\left(\Omega+A\phi\right)
	=0.
\end{eqnarray}
This equation can replace Eq. (\ref{scalar mode equation 3}), and we can solve scalar mode gravitational waves with Eqs. (\ref{scalar mode equation 1}), (\ref{scalar mode equation 4}), (\ref{scalar mode equation 5}), and (\ref{scalar mode equation new}).

Now, we study scalar mode gravitational waves in the following four cases.

\textbf{Case 1}: $\mathring{G}_{3,X}=1-2c_{2}\mathring{G}_{4,X}=0$. In this case, using Eq. (\ref{G4-G4XA2 neq 0}), we find that Eq. (\ref{scalar mode equation new}) requires $\Theta=0$. Thus Eq. (\ref{scalar mode equation 5}) {becomes}
\begin{eqnarray}
	\left(1-2c_{2}\mathring{G}_{4,X}\right)A\left(\Omega+2A\phi+\partial_{0}\Psi\right)
	=0.
\end{eqnarray}
It can be seen that the equation is an identity. We now have only two equations (\ref{scalar mode equation 1}) {and} (\ref{scalar mode equation 4}) left to constrain the three variables $\phi,\Psi,\Omega$, {which indicates that these variables can not be determined. So we do not consider this case.}

\textbf{Case 2}: $\mathring{G}_{3,X}=0,\!~1-2c_{2}\mathring{G}_{4,X} \neq 0$. In this case, using Eq. (\ref{G4-G4XA2 neq 0}), Eq. (\ref{scalar mode equation new}) also requires $\Theta=0$, while Eq. (\ref{scalar mode equation 5}) requires
\begin{eqnarray}
	\Omega+2A\phi+\partial_{0}\Psi=0.
\end{eqnarray}
{Then, from (\ref{scalar mode equation 4}) we have}
\begin{eqnarray}
	\label{case2 phi=0}
	\left(\mathring{G}_{4}-\mathring{G}_{4,X}A^2\right)\phi=0.
\end{eqnarray}
Using Eq. (\ref{G4-G4XA2 neq 0}), it requires $\phi=0$. Since $\Theta=\phi=0$, Eq. (\ref{Ri0j0 gauge invariant}) indicates that scalar mode gravitational waves are not allowed in this case.

\textbf{Case 3}: $\mathring{G}_{3,X} \neq 0,\!~1-2c_{2}\mathring{G}_{4,X}=0$. In this case, using Eq. (\ref{scalar mode equation 5}), we have $\Theta=0$. Then, Eq. (\ref{scalar mode equation new}) requires $\Omega+A\phi=0$. Using the above conditions, Eq. (\ref{scalar mode equation 1}) will require $\Psi=0$. Finally, from Eq. (\ref{scalar mode equation 4}), it can be seen that $\phi=0$. Since $\Theta=\phi=0$, {scalar mode gravitational waves are not allowed in this case.}

\textbf{Case 4}: $\mathring{G}_{3,X} \neq 0,\!~1-2c_{2}\mathring{G}_{4,X} \neq 0$. In this case, from Eq. (\ref{scalar mode equation new}) and (\ref{scalar mode equation 5}), we have
\begin{eqnarray}
	\label{case4 eq1}
	\Omega+A\phi &=&
	\frac{2\left(\mathring{G}_{4}-\mathring{G}_{4,X}A^2\right)}
         {\mathring{G}_{3,X}A^{2}}\partial_{0}\Theta,\\
	\label{case4 eq2}
	\Omega+2A\phi+\partial_{0}\Psi&=&
	-\frac{2\mathring{G}_{4}}{A\left(1-2c_{2}\mathring{G}_{4,X}\right)}\Theta.
\end{eqnarray}
{Substituting Eq. (\ref{case4 eq1}) from Eq. (\ref{case4 eq2}) yields}
\begin{eqnarray}
	\label{case4 eq3}
	A\phi+\partial_{0}\Psi=
	-\frac{2\mathring{G}_{4}}{A\left(1-2c_{2}\mathring{G}_{4,X}\right)}\Theta
	-\frac{2\left(\mathring{G}_{4}-\mathring{G}_{4,X}A^2\right)}{\mathring{G}_{3,X}A^{2}}\partial_{0}\Theta.
\end{eqnarray}
By substituting Eqs. (\ref{case4 eq1}) and (\ref{case4 eq2}) into Eq. (\ref{scalar mode equation 1}), we obtain
\begin{eqnarray}
	\label{case4 eq4}
	\Delta\Psi=
	-\frac{2\left(\mathring{G}_{4}-\mathring{G}_{4,X}A^2\right)}{\mathring{G}_{3,X}A^2}\Delta\Theta
	-\frac{3{\mathring{G}_{3,X}}^2 A^2+4\mathring{G}_{2,XX}\left(\mathring{G}_{4}-\mathring{G}_{4,X}A^2\right)}{2{\mathring{G}_{3,X}}^2 A} \partial_{0}\Theta.
\end{eqnarray}
From Eqs. (\ref{case4 eq3}) and (\ref{case4 eq4}), {it can be derived}
\begin{eqnarray}
	\label{case4 phi and theta}
	\Delta\phi=
	\frac{3{\mathring{G}_{3,X}}^2 A^2+4\mathring{G}_{2,XX}\left(\mathring{G}_{4}-\mathring{G}_{4,X}A^2\right)}{2{\mathring{G}_{3,X}}^2 A^2} \partial_{0}\partial_{0}\Theta
	-\frac{2\mathring{G}_{4}}{A^2 \left(1-2c_{2}\mathring{G}_{4,X}\right)}\Delta\Theta,
\end{eqnarray}
{which gives} the relationship between the amplitudes of the two scalar perturbations $\Theta$ and $\phi$ that affect scalar mode gravitational waves. {Finally, with Eqs. (\ref{case4 eq2}) and (\ref{case4 phi and theta}), we obtain the equation for the variable $\Theta$ from Eq. (\ref{scalar mode equation 4}):}
\begin{eqnarray}
	\label{scalar mode wave equation}
	 \mathcal{P}    \partial_{0}\partial_{0}\Theta
	-\mathcal{Q} 	\Delta\Theta
	=0.
\end{eqnarray}
where
\begin{eqnarray}
	\mathcal{P}  &=& {\left(1-2c_{2}\mathring{G}_{4,X}\right)}
	{\left(\mathring{G}_{4}-\mathring{G}_{4,X}A^2\right)
		} \mathcal{C}_1,  \\
	\mathcal{Q} &=& {{\mathring{G}_{3,X}}^2}\mathring{G}_{4}
	\mathcal{C}_2,\\
   \mathcal{C}_1&=&   3{\mathring{G}_{3,X}}^2 A^2
		   +4\mathring{G}_{2,XX}\left(\mathring{G}_{4}-\mathring{G}_{4,X}A^2\right),\\
   \mathcal{C}_2&=& {4{\mathring{G}_{4}} -\left(1-2c_{2}\mathring{G}_{4,X}\right)A^2}.
\end{eqnarray}

We can further divide Case 4 into following cases.

\textbf{Case 4.\!\!~1}: $\mathcal{C}_1=0,\!~ \mathcal{C}_2=0$. {In this case, we have
\begin{eqnarray}
	\Psi&=&
	-\frac{2\left(\mathring{G}_{4}-\mathring{G}_{4,X}A^2\right)}{\mathring{G}_{3,X}A^2}\Theta,\\
	\phi &=& \frac{1}{2}\Theta.
\end{eqnarray}
We also have $\mathcal{P}=\mathcal{Q}=0$ and so we can not solve the function  $\Theta$ with Eq. (\ref{scalar mode wave equation}). Therefore, we do not consider this nonphysical case.}

\textbf{Case 4.\!\!~2}: $\mathcal{C}_1 \neq 0,\!~ \mathcal{C}_2=0$. In this case, Eq. (\ref{scalar mode wave equation}) becomes $\partial_{0}\partial_{0}\Theta=0$. This requires $\Theta=0$, and then by Eq. (\ref{case4 phi and theta}), $\phi$ is also zero. Thus, generalized Proca theory does not allow scalar mode gravitational waves in this case.

\textbf{Case 4.\!\!~3}: $\mathcal{C}_1 = 0,\!~ \mathcal{C}_2 \neq 0$. {In this case, we have $\mathcal{P}=0$ and $\mathcal{Q}\neq 0$, which indicates that} Eq. (\ref{scalar mode wave equation}) becomes $\Delta\Theta=0$. This requires $\Theta=0$. And then by Eq. (\ref{case4 phi and theta}), $\phi$ is also zero. Therefore, generalized Proca theory  does not allow scalar mode gravitational waves in this case {either}.

\textbf{Case 4.\!\!~4}: $\mathcal{C}_1 \neq 0,\!~ \mathcal{C}_2 \neq 0$. In this case, Eq. (\ref{scalar mode wave equation}) is a wave equation and generalized Proca theory allows scalar mode gravitational waves with one degree of freedom. The speed of the scalar mode gravitational waves $c_{S}$ is
\begin{eqnarray}
	\label{cS}
    c_{S}^2=
	\frac{\mathcal{Q}}
	{\mathcal{P}}.
\end{eqnarray}
To ensure the linear stability of the solution, we require $c_{S}^2 \textgreater 0$.

Finally, we consider a plane wave solution of $\Theta$ satisfying Eq. (\ref{scalar mode wave equation}) with a propagation direction of $+z$:
\begin{eqnarray}
	\label{plane wave theta}
	\Theta=\Theta_{0}~\!e^{ikx},\quad \frac{k^{0}}{k^{3}}=c_{S}.
\end{eqnarray}
By substituting (\ref{plane wave theta}) into Eq. (\ref{case4 phi and theta}), we obtain
\begin{eqnarray}
	\label{delta phi =}
	\Delta\phi=
	\left[
	-\frac{\mathcal{C}_1}{2{\mathring{G}_{3,X}}^2 A^2} k_{0}^2
	+\frac{2\mathring{G}_{4}}{A^2 \left(1-2c_{2}\mathring{G}_{4,X}\right)} k_{3}^{2}
	\right]
	\Theta_{0}~\!e^{ikx}.
\end{eqnarray}
Therefore, $\phi$ has a plane wave solution $\phi=\phi_{0}~\!e^{ikx}$ with
\begin{eqnarray}
	\label{plane wave phi}
	\phi_{0}
	=\left[
	\frac{\mathcal{C}_1}{2{\mathring{G}_{3,X}}^2 A^2} c_{S}^2
	-\frac{2\mathring{G}_{4}}{A^2 \left(1-2c_{2}\mathring{G}_{4,X}\right)}
	\right]
	\Theta_{0}.
\end{eqnarray}
Using Eq. (\ref{P1-P6 gauge invariant}), we have
\begin{eqnarray}
	P_{1}&=&
	\left[
    \frac{{\mathring{G}_{3,X}}^2 A^2 -\mathcal{C}_1}
         {2{\mathring{G}_{3,X}}^2 A^2}
    k_{0}^2
	+\frac{2\mathring{G}_{4}}{A^2 \left(1-2c_{2}\mathring{G}_{4,X}\right)} k_{3}^{2}
	\right]
	\Theta_{0}~\!e^{ikx};
	 \\
	P_{6}&=& \frac{1}{2} k_{0}^2 \Theta_{0}~\!e^{ikx}.
\end{eqnarray}
{So the amplitude ratio $\mathcal{R}$ of the longitudinal mode to the breathing mode is}
\begin{eqnarray}
	\label{ratio P1,P6}
	\mathcal{R}=
	\left|\frac{P_1}{P_6}\right|=
	\left|
	\frac{{\mathring{G}_{3,X}}^2 A^2 -\mathcal{C}_1}
         {{\mathring{G}_{3,X}}^2 A^2}
	+\frac{4\mathring{G}_{4}}{A^2 \left(1-2c_{2}\mathring{G}_{4,X}\right){c_{S}}^{2}}
	\right|.
\end{eqnarray}
It can be seen that the scalar mode of generalized Proca theory is generally a mixture mode of the breathing and longitudinal modes. However, when $\mathcal{R}=0$, the scalar mode is just a pure breathing mode. It should be pointed out that $\mathcal{R}$ only depends on theoretical parameters. Therefore, once we provide the specific forms of $G_{2}, G_{3}, G_{4}, G_{5}, c_{2}, d_{2}$ in the action and the background vector field, we can uniquely determine the value of $\mathcal{R}$. Whether it is $0$ clearly tells us whether the scalar mode gravitational wave is a breathing mode or a mixture of the longitudinal mode and the breathing mode. Furthermore, it should be pointed out that the condition $\mathcal{R}=0$ is not equivalent to $c_{S}=1$.

Based on the above research, we find that the existence of scalar mode gravitational waves in generalized Proca theory depends on the parameter spaces. We summarize the results in Table \ref{tab:scalar}.

\begin{center}
	\begin{table}[htbp]
			\begin{tabular}{|c|c|c|}
				\hline\hline
				\textbf{Cases} & \textbf{Conditions} &  \textbf{Scalar DoF} \\
				\hline
				case 1 & $\mathring{G}_{3,X}=1-2c_{2}\mathring{G}_{4,X}=0$. & - \\
				\hline
				case 2 &$\mathring{G}_{3,X}=0,\!~1-2c_{2}\mathring{G}_{4,X} \neq 0$. & 0 \\
				\hline
				case 3 &$\mathring{G}_{3,X} \neq 0,\!~1-2c_{2}\mathring{G}_{4,X}=0$. & 0 \\
				\hline
				case 4.\!\!~1 & $\mathring{G}_{3,X} \neq 0,
				\!~1-2c_{2}\mathring{G}_{4,X} \neq 0,\!~
				\mathcal{C}_1=0,
				\!~ \mathcal{C}_2=0$. & - \\
				\hline
				case 4.\!\!~2 & $\mathring{G}_{3,X} \neq 0,
				\!~1-2c_{2}\mathring{G}_{4,X} \neq 0,\!~
				\mathcal{C}_1 \neq 0,
				\!~ \mathcal{C}_2=0$. & 0 \\
				\hline
				case 4.\!\!~3 & $\mathring{G}_{3,X} \neq 0,
				\!~1-2c_{2}\mathring{G}_{4,X} \neq 0,\!~
				\mathcal{C}_1=0,
				\!~ \mathcal{C}_2 \neq 0$. & 0 \\
				\hline
				case 4.\!\!~4 & $\mathring{G}_{3,X} \neq 0,
				\!~1-2c_{2}\mathring{G}_{4,X} \neq 0,\!~
				\mathcal{C} \neq 0,
				\!~ \mathcal{C}_2 \neq 0$. & 1 \\
				\hline\hline
		\end{tabular}
		\caption{The number of scalar mode gravitational waves in various parameter spaces. The number in the rightmost column of the table represents the degrees of freedom of the scalar mode gravitational waves in the corresponding case. The notation "-" indicates that {the parameter space is not physical}.}
		\label{tab:scalar}
	\end{table}
\end{center}

\section{Conclusion}
\label{sec: 6}
In this paper, under the background of the homogeneous and isotropic Minkowski space-time, we study the polarization modes of gravitational waves in generalized Proca theory under linearized perturbations. We first obtained the background equations and the linear perturbation equations. Then we analyzed the polarization modes of gravitational waves and discussed their linear stability.

We found that the polarization modes of gravitational waves in generalized Proca theory depend on the parameter spaces. There are different vector and scalar polarization modes in different parameter spaces. The specific results have been summarized in Table {\ref{tab:vector}} and Table {\ref{tab:scalar}}, respectively. Generalized Proca theory allows at most two tensor modes, two vector modes and one scalar mode. The scalar mode is typically a mixture mode of the breathing and longitudinal modes. However, when $\mathcal{R}$ in Eq. (\ref{ratio P1,P6}) is zero, the scalar mode is just a pure breathing mode.

We also found that the amplitude of vector mode gravitational waves satisfies the condition (\ref{Xi<<cT}). This allows us to give the upper limit of the amplitude of vector mode gravitational waves by detecting the speed of tensor mode gravitational waves. Combining the results of GW170817, we found that the amplitude of vector gravitational waves has an upper bound of $|h| \ll 10^{-15}$. It should be noted that as the speed of tensor gravitational waves approaches the speed of light, the upper bound of the amplitude of vector gravitational waves decreases. Specifically, if the speed of tensor modes is strictly equal to the speed of light, then the amplitude of vector modes is zero. The relationship between the amplitude of vector modes and the speed of tensor modes is an interesting phenomenon. Does this phenomenon also exist in other vector-tensor gravity? Could we find a physical explanation for this phenomenon? This requires further research.

Finally, we need to point out that we only considered the case of the homogeneous and isotropic Minkowski background. At this point, the background vector $\mathring{A}_{a}$ has only a nonvanishing temporal component. In the case of a small anisotropic background, the background vector $\mathring{A}_{a}$ also has nonvanishing small spatial components. In this case, the background vector is time-like. We can always use a Lorentz transformation to make the background vector have only nonvanishing temporal component without losing generality. It can be seen that the polarization modes of gravitational waves in the anisotropic case can be directly obtained by using the Lorentz transformation on the polarization modes of gravitational waves in the isotropic case we have studied. For a general discussion of the transformation law of the polarization modes of gravitational waves under the Lorentz transformation, one can refer to Ref. \cite{Y.Liu}.

The polarization modes of gravitational waves in generalized Proca theory can be divided into quite rich cases by parameter spaces. The appropriate parameter spaces can be expected to be selected in the detection of gravitational wave polarization modes by Lisa, Taiji and TianQin \cite{lisa,taiji,tianqin} in the future.

\section*{Acknowledgments}
{We would like to thank Lavinia Heisenberg for useful discussion.} This work is supported in part by the National Key Research and Development Program of China (Grant No. 2020YFC2201503), the National Natural Science Foundation of China (Grants No. 11875151 and No. 12247101), the 111 Project (Grant No. B20063), the Department of education of Gansu Province: Outstanding Graduate ``Innovation Star" Project (Grant No. 2022CXZX-059), the Major Science and Technology Projects of Gansu Province,  and ``Lanzhou City's scientic research funding subsidy to Lanzhou University".

\appendix
\section{Linear perturbation equation}
\label{app: A}

For Eq. (\ref{vector perturbations equation}), the specific expression of $\mathcal{V}_{a}$ is given by
\begin{eqnarray}
	\mathcal{V}_{a}
	&=&2\Box B_{a}
	-2\mathring{A}^{b}\partial_{a}\partial_{c}h^{c}_{~b}
	+2\mathring{A}^{b}\Box h_{ab}
	-2\mathring{G}_{2,X}B_{a}
	-2\mathring{G}_{2,X}\mathring{A}^{b}h_{ab}
	\nonumber \\
	&+&2\mathring{G}_{3,X}\mathring{A}^{b}\partial_{a}B_{b}
	+\mathring{G}_{3,X}\mathring{A}^{b}\mathring{A}^{c}\partial_{a}h_{bc}
	-2\mathring{G}_{3,X}\mathring{A}_{a}\partial_{b}B^{b}
	-\mathring{G}_{3,X}\mathring{A}_{a}\mathring{A}^{b}\partial_{b}h
	\nonumber \\
	&-&4\mathring{G}_{4,X}\partial_{a}\partial_{b}B^{b}
	-2\mathring{G}_{4,X}\mathring{A}^{b}\partial_{a}\partial_{b}h
	-4c_{2}\mathring{G}_{4,X}\Box B_{a}
	+2\mathring{G}_{4,X}\mathring{A}^{b}\partial_{a}\partial_{c}h_{b}^{~c}
	\nonumber \\
	&+&4c_{2}\mathring{G}_{4,X}\mathring{A}^{b}\partial_{a}\partial_{c}h_{b}^{~c}
	+2\mathring{G}_{4,X}\mathring{A}^{b}\partial_{b}\partial_{c}h_{a}^{~c}
	-2\mathring{G}_{4,X}\mathring{A}_{a}\partial_{b}\partial_{c}h^{bc}
	-2\mathring{G}_{4,X}\mathring{A}^{b}\Box h_{ab}
	\nonumber \\
	&-&4c_{2}\mathring{G}_{4,X}\mathring{A}^{b}\Box h_{ab}
	+2\mathring{G}_{4,X}\mathring{A}_{a}\Box h
	+\left[-2+4\left(1+c_{2}\right)\mathring{G}_{4,X}\right]\partial_{a}\partial_{b}B^{b}
	\nonumber \\
	&+&2\mathring{G}_{2,XX}\mathring{A}_{a}\mathring{A}^{b}B_{b}
	+\mathring{G}_{2,XX}\mathring{A}_{a}\mathring{A}^{b}\mathring{A}^{c}h_{bc}.
\end{eqnarray}

For Eq. (\ref{tensor perturbations equation}), the specific expression of $\mathcal{T}_{ab}$ is
\begin{eqnarray}
	\mathcal{T}_{ab}
	&=&-2\mathring{G}_{2}h_{ab}
	-2\mathring{A}_{b}\Box B_{a}
	-2\mathring{A}_{a}\Box B_{b}
	+2\mathring{A}_{b}\partial_{a}\partial_{c}B^{c}
	+2\mathring{A}_{a}\partial_{b}\partial_{c}B^{c}
	\nonumber \\
	&+&2\mathring{A}_{b}\mathring{A}^{c}\partial_{a}\partial_{d}h_{c}^{~d}
	+2\mathring{A}_{a}\mathring{A}^{c}\partial_{b}\partial_{d}h_{c}^{~d}
	-2\mathring{A}_{b}\mathring{A}^{c}\Box h_{ac}
	-2\mathring{A}_{a}\mathring{A}^{c}\Box h_{bc}
	\nonumber \\
	&-&2\mathring{G}_{4}
	\left(
	\partial_{a}\partial_{b}h
	-\partial_{a}\partial_{c}h^{c}_{~b}
	-\partial_{b}\partial_{c}h^{c}_{~a}
	+\Box h_{ab}
	+\eta_{ab} \partial_{c}\partial_{d}h^{cd}
	-\eta_{ab} \Box h
	\right)
	\nonumber \\
	&+&2\mathring{G}_{2,X}\mathring{A}_{b}B_{a}
	+2\mathring{G}_{2,X}\mathring{A}_{a}B_{b}
	+2\mathring{G}_{2,X}\eta_{ab}\mathring{A}^{c}B_{c}
	+2\mathring{G}_{2,X}\mathring{A}_{b}\mathring{A}^{c}h_{ac}
	\nonumber \\
	&+&2\mathring{G}_{2,X}\mathring{A}_{a}\mathring{A}^{c}h_{bc}
	+\mathring{G}_{2,X}\eta_{ab}\mathring{A}^{c}\mathring{A}^{d}h_{cd}
	+2\mathring{G}_{3,X}\mathring{A}_{a}\mathring{A}_{b}\partial_{c}B^{c}	
	\nonumber \\
	&+&\mathring{G}_{3,X}\mathring{A}_{a}\mathring{A}_{b}\mathring{A}^{c}\partial_{c}h
	-2\mathring{G}_{3,X}\eta_{ab}\mathring{A}^{c}\mathring{A}^{d}\partial_{d}B_{c}
	-\mathring{G}_{3,X}\eta_{ab}\mathring{A}^{c}\mathring{A}^{d}\mathring{A}^{e}\partial_{e}h_{cd}
	\nonumber \\
	&+&2\mathring{G}_{4,X}\mathring{A}^{c}\partial_{a}\partial_{b}B_{c}
	+\mathring{G}_{4,X}\mathring{A}^{c}\mathring{A}^{d}\partial_{a}\partial_{b}h_{cd}
	+2\mathring{G}_{4,X}\mathring{A}^{c}\partial_{a}\partial_{b}B_{c}
	\nonumber \\
	&+&\mathring{G}_{4,X}\mathring{A}^{c}\mathring{A}^{d}\partial_{a}\partial_{b}h_{cd}
	-2\mathring{G}_{4,X}\mathring{A}^{c}\partial_{c}\partial_{a}B_{b}
	-2\mathring{G}_{4,X}\mathring{A}_{b}\partial_{a}\partial_{c}B^{c}
	\nonumber \\
	&-&4c_{2}\mathring{G}_{4,X}\mathring{A}_{b}\partial_{a}\partial_{c}B^{c}
	-2\mathring{G}_{4,X}\mathring{A}^{c}\partial_{c}\partial_{b}B_{a}
	-2\mathring{G}_{4,X}\mathring{A}_{a}\partial_{b}\partial_{c}B^{c}
	\nonumber \\
	&-&4c_{2}\mathring{G}_{4,X}\mathring{A}_{a}\partial_{b}\partial_{c}B^{c}
	+2\mathring{G}_{4,X}\mathring{A}_{b}\Box B_{a}
	+4c_{2}\mathring{G}_{4,X}\mathring{A}_{b}\Box B_{a}
	\nonumber \\
	&+&2\mathring{G}_{4,X}\mathring{A}_{a}\Box B_{b}
	+4c_{2}\mathring{G}_{4,X}\mathring{A}_{a}\Box B_{b}
	+4\mathring{G}_{4,X}\eta_{ab}\mathring{A}^{c}\partial_{c}\partial_{d}B^{d}
	\nonumber \\
	&-&2\mathring{G}_{4,X}\mathring{A}_{b}\mathring{A}^{c}\partial_{a}\partial_{d}h_{c}^{~d}
	-4c_{2}\mathring{G}_{4,X}\mathring{A}_{b}\mathring{A}^{c}\partial_{a}\partial_{d}h_{c}^{~d}
	-2\mathring{G}_{4,X}\mathring{A}_{a}\mathring{A}^{c}\partial_{b}\partial_{d}h_{c}^{~d}
	\nonumber \\
	&-&4c_{2}\mathring{G}_{4,X}\mathring{A}_{a}\mathring{A}^{c}\partial_{b}\partial_{d}h_{c}^{~d}
	-2\mathring{G}_{4,X}\mathring{A}^{c}\mathring{A}^{d}\partial_{c}\partial_{d}h_{ab}
	+2\mathring{G}_{4,X}\mathring{A}_{a}\mathring{A}_{b}\partial_{c}\partial_{d}h^{cd}
	\nonumber \\
	&+&2\mathring{G}_{4,X}\eta_{ab}\mathring{A}^{c}\mathring{A}^{d}\partial_{c}\partial_{d}h
	-4G_{4,X}\eta_{ab}\mathring{A}^{c}\Box B_{c}
	+2\mathring{G}_{4,X}\mathring{A}_{b}\mathring{A}^{c}\Box h_{ac}
	\nonumber \\
	&+&4c_{2}\mathring{G}_{4,X}\mathring{A}_{b}\mathring{A}^{c}\Box h_{ac}
	+2\mathring{G}_{4,X}\mathring{A}_{a}\mathring{A}^{c}\Box h_{bc}
	+4c_{2}\mathring{G}_{4,X}\mathring{A}_{a}\mathring{A}^{c}\Box h_{bc}
	\nonumber \\
	&-&2\mathring{G}_{4,X}\mathring{A}_{a}\mathring{A}_{b}\Box h
	-2\mathring{G}_{4,X}\eta_{ab}\mathring{A}^{c}\mathring{A}^{d}\Box h_{cd}
	-2\mathring{G}_{2,XX}\mathring{A}_{a}\mathring{A}_{b}\mathring{A}^{c}B_{c}
	\nonumber \\
	&-&\mathring{G}_{2,XX}\mathring{A}_{a}\mathring{A}_{b}\mathring{A}^{c}\mathring{A}^{d}h_{cd}.
\end{eqnarray}


\end{document}